\documentclass[aps,pra,twocolumn,superscriptaddress]{revtex4-2}
\usepackage[T1]{fontenc}
\usepackage{soul}
\usepackage{textcomp}
\usepackage[english]{babel}
\usepackage{graphicx}
\usepackage{dcolumn}
\usepackage{bm}
\usepackage{bm}
\usepackage{amsmath}
\usepackage{verbatim}     
\usepackage[dvipsnames]{xcolor}       
\usepackage[normalem]{ulem}   
\usepackage{bbold}
\usepackage{mathrsfs}
\usepackage[mathscr]{eucal}
\usepackage{physics}
\usepackage{braket}
\usepackage{amsmath}
\usepackage{mathrsfs}

\DeclareUnicodeCharacter{0308}{\"}
\newcommand{\qo}[1]{``#1''}                               		
\newcommand{\beq}{\begin{equation}}
\newcommand{\eeq}{\end{equation}}
\newcommand{\bei}{\begin{itemize}}			
\newcommand{\eei}{\end{itemize}}			

\newcommand{\Cite}{\unskip~\cite}
\newcommand{\Eqref}{\unskip~\eqref}
\newcommand{\figref}{\unskip~\ref}

\usepackage{hyperref}
\hypersetup{
   pdfpagemode=None,
   pdfstartpage=1,
   pdfmenubar=true,
   pdftoolbar=true,
   colorlinks = true,
   linkcolor=blue,
   citecolor=blue,
   urlcolor=blue,
   bookmarksopen=false
 }

\begin{document}

\title{Skyrmionic polarization textures in structured dielectric planar media}

\author{Francesco Di Colandrea}
\email{francesco.dicolandrea@unina.it}
\affiliation{Dipartimento di Fisica \qo{Ettore Pancini}, Universit\`{a} degli Studi di Napoli Federico II, Complesso Universitario di Monte Sant'Angelo, Via Cintia, 80126 Napoli, Italy}
\author{Lorenzo Marrucci}
\affiliation{Dipartimento di Fisica \qo{Ettore Pancini}, Universit\`{a} degli Studi di Napoli Federico II, Complesso Universitario di Monte Sant'Angelo, Via Cintia, 80126 Napoli, Italy}
\affiliation{CNR-ISASI, Institute of Applied Science and Intelligent Systems, Via Campi Flegrei 34, 80078 Pozzuoli (NA), Italy}
\author{Filippo Cardano}
\email{filippo.cardano2@unina.it}
\affiliation{Dipartimento di Fisica \qo{Ettore Pancini}, Universit\`{a} degli Studi di Napoli Federico II, Complesso Universitario di Monte Sant'Angelo, Via Cintia, 80126 Napoli, Italy}

\begin{abstract}
Skyrmionic patterns of optical fields have recently emerged across diverse photonic platforms. Here, we show that such textures also arise in the polarization eigenstates of light propagation through flat dielectric devices with an engineered, space-dependent optic-axis orientation. We focus on two-dimensional periodic structures, where propagation through multiple devices maps onto quantum dynamics on a synthetic optical lattice. Adopting the condensed-matter framework, a spatial period defines an effective Brillouin zone, and polarization eigenstates can be grouped in two bands, with the role of energy played by the opposite phase delay. When such eigenstates exhibit skyrmionic textures, the corresponding lattice model shows the topology of a Chern insulator. These structures result from the interaction between the optical field and the medium and do not reflect a topological structure of the medium itself.
We validate these concepts in a system of three tunable liquid-crystal metasurfaces. Using quantum process tomography based on supervised machine learning, we reconstruct the polarization eigenmodes over one spatial period.  
We identify configurations of the devices' parameters that lead to topologically non-trivial bands, where we directly observe skyrmionic eigenpolarization textures. Along the analogy with condensed matter, we also extract local observables of lattice models, such as the Berry curvature and the quantum metric. We finally report a numerical simulation of an all-optical quantum Hall effect emerging when light propagates through a sequence of such devices, arranged so as to mimic the effect of an external force on the lattice.

\end{abstract}
\maketitle

\section{Introduction}
Originally introduced in particle physics \Cite{SKYRME1962556}, skyrmions are topologically stable configurations of vector fields that have been observed in different scenarios, ranging from magnetic materials \Cite{doi:10.1126/science.1166767,Finocchio_2016,Fert2017,Bogdanov2020,Tokura2021,Han2022} to liquid crystals \Cite{PhysRevLett.126.047801}, Bose-Einstein condensates \Cite{AlKhawaja2001}, twistronics \Cite{PhysRevX.12.031020}, and acoustics \Cite{PhysRevLett.127.144502}. The topological stability of these structures makes them ideal candidates for high-density data storage and transfer \Cite{liu2016skyrmions,Fert2017,Zhang_2020,Han2022,LimaFernandes2022}. When used as carriers of information, the latter is encoded in their topological invariant, the Skyrme number, indicating the number of times the vector field wraps around a unit sphere. In the context of condensed matter, the Skyrme number is usually referred to as the Chern number \Cite{RevModPhys.82.3045,asbóth2016short}. 
\begin{figure*}[t!]
    \centering
    \includegraphics[width=\linewidth]{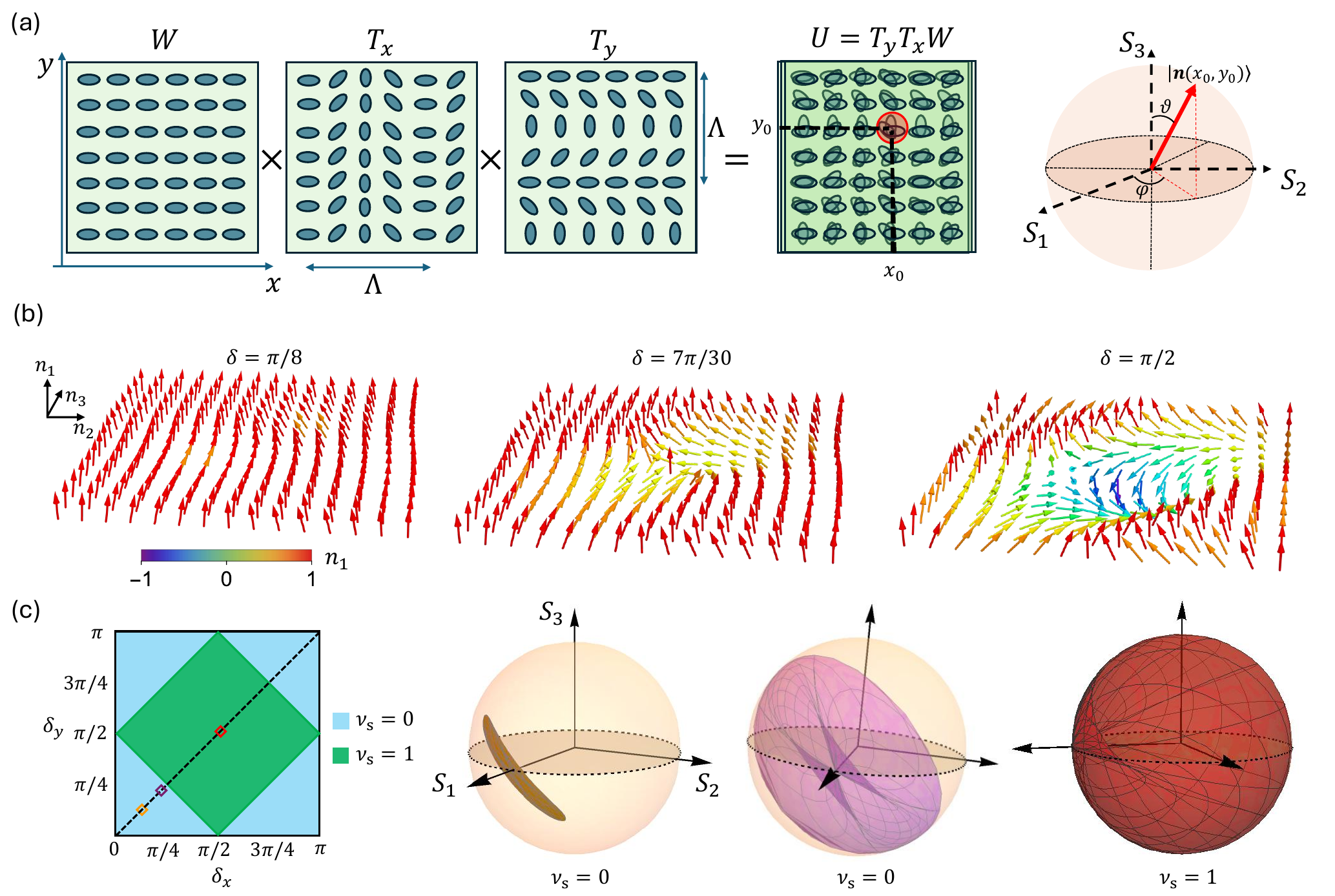}
    \caption{\textbf{Eigenpolarization of structured planar media.} (a) Three adjacent liquid-crystal metasurfaces define a 2D optical operator $U$, implementing a complex polarization transformation. Its action is periodic across a characteristic distance $\Lambda$, which defines a Brillouin zone (BZ) in real space (see Sec. \figref{sec:bloch} for more details). Accordingly, each transverse position on the metasurfaces' plane identifies a quasi-momentum value according to the mapping $\textbf{q}=-2\pi\textbf{{r}}/\Lambda$. Each local polarization transformation $U(x_0,y_0)$ thus corresponds to the Bloch-diagonal form of a discrete lattice evolution operator at a given quasi-momentum. The associated polarization eigenstate, $\ket{\textbf{n}(x_0,y_0)}$, can be visualized on the Bloch-Poincaré sphere, with the components of $\textbf{n}$ giving the Stokes parameters $\textbf{S}=(S_1,S_2,S_3)$. (b) Eigenpolarization textures for three representative cases along the curve $\delta_x=\delta_y\equiv\delta$. Arrows give the orientation of the local eigenpolarization on the sphere. The patterns preserve the orientation in the cases ${\delta=\pi/8}$ and ${\delta=7\pi/30}$, while a skyrmion appears at ${\delta=\pi/2}$, where the eigenpolarization flips at the center. (c)~Topological phase diagram for the operator ${U=T_y(\delta_y)T_x(\delta_x)W}$ as a function of $\delta_x$ and $\delta_y$. For ${\delta=\pi/8}$ (orange) and ${\delta=7\pi/30}$ (purple), the eigenpolarization field $\textbf{n}(\textbf{q})$ does not fully cover the Bloch sphere, which corresponds to a vanishing Skyrme number (${\nu_\text{s}=0}$). In contrast, for ${\delta=\pi/2}$ (red), the Bloch sphere is completely covered as $\textbf{q}$ varies across the BZ, indicating a topologically non-trivial phase (${\nu_\text{s}=1}$).}
    \label{fig:fig1}
\end{figure*}

Only recently have skyrmions been observed in optical fields \Cite{Shen2024,Yang_2025}, first in the evanescent electric field of surface plasmon polariton (SPP) waves in a hexagon-shaped resonator \Cite{doi:10.1126/science.aau0227} and soon after as spin-skyrmions in evanescent SPP fields carrying orbital angular momentum \Cite{Du2019}. Since then, skyrmions have garnered growing interest in photonics, leading to several demonstrations of optical skyrmions or skyrmion-like structures. For example, SPP fields have also been tailored to create more exotic structures, known as merons, corresponding to field textures with a fractional topological invariant \Cite{Dai2020,Xiong2021}. Skyrmions as topological realizations of full Poincaré beams have also been proposed and observed \Cite{Beckley:10,supervortices,PhysRevA.102.053513,Shen:21}. Further demonstrations have been reported in microcavities filled with chiral liquid crystals \Cite{Krol:21}, nonlinear media \Cite{Karnieli2021}, propagation through structured birefringent materials \Cite{PhysRevLett.134.083802,PhysRevApplied.22.054038} or spatial light modulators \Cite{Sugic2021,Shen2022}, metafibers \Cite{He2024}, and spatio-temporal toroidal pulses \Cite{Shen2021}. A recent work has also extended the notion of skyrmions to the quantum regime \Cite{Ornelas2024}, demonstrating a non-local topological structure that emerges in the hybrid-entangled two-photon correlation function. The ultrasmall size of typical skyrmionic structures, combined with their topological stability, has evoked their potential use in microscopy, sensing, and robust all-optical information processing \Cite{Shen2024}. On the other hand, some research has revealed the fragility of skyrmionic structures in common experimental scenarios, underscoring the importance of always testing their robustness without assuming it \Cite{Chen2025}.

In this work, we report the observation of optical skyrmions that do not result from spatially structuring the optical field, but instead arise in the polarization eigenstates of spatially structured planar media. Such skyrmionic eigenpolarization textures thus emerge as intrinsic properties of the probed material and of its interaction with the optical field rather than in the field structure. A realization of this concept has recently been demonstrated in an optical Raman lattice of ultracold atoms by resolving polarization eigenfunctions via spin-resolved time-of-flight imaging \Cite{PhysRevResearch.5.L032016}. Here, we consider a sequence of three subsequent liquid-crystal metasurfaces with negligible optical propagation distance, arranged in a two-dimensional (2D) configuration, with periodic modulation of the optic axes and electrically tunable birefringence \Cite{Rubano:19}. The so-built optical operator is shown to be in a one-to-one mapping with the single-step evolution operator of a quantum walk over a discrete 2D lattice, as first shown in Ref. \Cite{DErrico:20}. Under specific birefringence settings, the eigenpolarization texture of this setup exhibits a skyrmionic topology that is revealed via quantum process tomography.

In particular, a deep neural network is trained to deliver a fast reconstruction of the process eigenstructure from a minimal set of six polarimetric measurements \Cite{DiColandrea:23}. From the reconstructed 2D maps, we compute the Skyrme number to characterize the topological nature of such structures. From the same patterns, we also extract the real and imaginary parts of the quantum geometric tensor \Cite{Provost1980}, namely the quantum metric and the Berry curvature. These are local observables that represent key concepts in the description of the geometry of the parameter space of the simulated lattice model \Cite{Berry1989,PhysRevLett.131.240001,metricexp}, and have been observed to play a pivotal role in two-band wavepacket dynamics in the presence of external driving \Cite{PhysRevLett.111.135302,PhysRevB.93.245113,PhysRevLett.121.020401,PhysRevB.97.195422,Gianfrate2020}. Finally, we discuss how this non-trivial topological properties can be experimentally revealed by an optical process mimicking the quantum Hall effect \Cite{DErrico:20}.

\section{Theory}
\subsection{Optical eigenpolarizations}
A liquid-crystal metasurface can be modeled as a waveplate with a patterned optic-axis orientation and uniform yet tunable birefringence \Cite{Rubano:19}. Optical polarizations aligned either with the ordinary or the extraordinary directions accumulate a phase delay $\delta_o=n_o k d$ and $\delta_e=n_e k d$, respectively, where $n_o$ and $n_e$  are the ordinary and extraordinary refractive indices, ${k=2\pi/\lambda}$ is the longitudinal wavevector, and $d$ is the plate thickness. The index $n_e$ can be tuned by applying an electric field perpendicular to the cell. In the basis of circular polarizations, where ${\ket{L}=\left(1,0 \right)^T}$ and ${\ket{R}=\left(0,1 \right)^T}$ are the left-handed and right-handed circular polarization states ($T$ stands for the transpose operator), the associated Jones matrix reads
\begin{equation}
L(\delta,\theta\left(x,y\right))=\begin{pmatrix}
\cos{\dfrac{\delta}{2}} &i\sin{\dfrac{\delta}{2}e^{-2i\theta(x,y)}}\\[5pt]
i\sin{\dfrac{\delta}{2}e^{2i\theta(x,y)}} & \cos{\dfrac{\delta}{2}}
\end{pmatrix},
\end{equation}
where ${\delta=\delta_e-\delta_o}$ is the birefringence parameter and $\theta(x,y)$ is the space-dependent optic-axis orientation, which can be prepared arbitrarily \Cite{Rubano:19}. We are omitting here a global phase factor $(\delta_e+\delta_o)/2$, which does not play any role for our purposes. 

In this work, we consider the optical operator $U$ describing the actions of three closely stacked liquid-crystal metasurfaces, defined as:
\begin{equation}
    U(x,y) = T_y(\delta_y)T_x(\delta_x)W,
    \label{eqn:opticalU}
\end{equation}
where 
\begin{equation}
    W=L(\pi/2,0)=\dfrac{1}{\sqrt{2}}\begin{pmatrix}
1 &i\\
i &1
    \end{pmatrix}
\end{equation}
is a metasurface with fixed birefringence and uniform optic axis, while $T_x$ ($T_y$) is a polarization grating ($g$-plate \Cite{DErrico:20}), characterized by a linear modulation of the optic axis along $x$ ($y$): $\theta_x(x)=\pi x/\Lambda$ ($\theta_y(y)=\pi y/\Lambda$), where $\Lambda$ represents the spatial period of the liquid-crystal molecular director, as shown in Fig. \figref{fig:fig1}(a). 

The operator $U$ thus implements a space-dependent polarization transformation that is periodic over a distance $\Lambda$, both along the $x$ and the $y$ directions. By restricting our analysis to sufficiently small spatial regions, we can approximate the metasurfaces as exhibiting a locally uniform optic-axis orientation. Accordingly, for a given setting of the $\delta_x$ and $\delta_y$ parameters, if we prepare a small beam, with a transverse size much smaller than $\Lambda$ (ideally, a plane wave in Fourier space), centered in $(x_0,y_0)$ and polarized along one of the local polarization eigenstates of the metasurfaces, say ${\ket{\textbf{n}(x_0,y_0)}}$, where $\textbf{n}$ is the corresponding unit vector on the Poincaré sphere for that polarization (see Fig. \figref{fig:fig1}(a)), the resulting effect will only be a global phase factor acquired by the input beam:
\begin{equation}
U\ket{\textbf{n}(x_0,y_0)}\otimes\ket{x_0,y_0}=e^{i\phi(x_0,y_0)}\ket{\textbf{n}(x_0,y_0)}\otimes\ket{x_0,y_0}.    
\label{eqn:phase-energy}
\end{equation}
Equation \Eqref{eqn:phase-energy} reads as an eigenvalue equation. As a consequence of the local unitarity of the transformation, the orthogonal eigenpolarization, oriented along the opposite direction on the Poincaré sphere, $-\textbf{n}$, will simply acquire the conjugate phase factor: ${\ket{-\textbf{n}}\rightarrow e^{-i\phi}\ket{-\textbf{n}}}$.
This suggests that the optical operator features a spectral decomposition of the form
\begin{equation}
\begin{split}
&U=\int_{0}^{\Lambda}\!\!\!\int_{0}^{\Lambda} \text{d}x\text{d}y\, \ketbra{x,y}\otimes  \\&(e^{i\phi(x,y)}\ketbra{\textbf{n}(x,y)}
+e^{-i\phi(x,y)}\ketbra{-\textbf{n}(x,y)}).    
\end{split}
\end{equation}

The discussion above clearly holds only for the case of localized input beams, where the eigenpolarizations can be considered to be approximately uniform. The variation of the local eigenpolarization across a full spatial period of the metasurfaces encodes the topological properties of the global eigenpolarization texture. Figure \figref{fig:fig1}(b) shows the eigenpolarization texture for three different cases along the ${\delta_x=\delta_y\equiv\delta}$ direction: ${\delta=\pi/8,7\pi/30,\pi/2}$. Here, the arrows give the orientation of the polarization eigenstates at each position, and their color is associated with the value of the $n_1$ component. For ${\delta=\pi/8}$ and ${\delta=7\pi/30}$, the eigenpolarizations vary smoothly over the spatial period and always preserve the positive orientation along one axis. In other words, the pattern can be continuously deformed into a uniform polarization state, which indicates a trivial topological texture. In the case ${\delta=\pi/2}$, we notice that the central vector is inverted with respect to the background. This feature is the hallmark of a skyrmion. 

Such an inversion gives rise to a quantized topological charge. The topology of the eigenpolarization pattern is quantified by the Skyrme number \Cite{Shen2024},
\begin{equation}
\nu_\text{s}=\frac{1}{4\pi}\int_{0}^{\Lambda}\!\!\!\int_{0}^{\Lambda}\text{d}{x}\text{d}{y}\,\textbf{n}(x,y)\cdot \left(\frac{\partial \textbf{n}}{\partial x}\times \frac{\partial \textbf{n}}{\partial y} \right),
\label{eqn:skyrme}
\end{equation}
counting the number of times the eigenpolarizations cover the Poincaré sphere as the transverse position varies across one spatial period.
In our setup, the Skyrme number depends on the birefringence settings of the $g$-plates along $x$ and $y$:
\begin{equation}
\nu_\text{s}(\delta_x,\delta_y) =
\Theta\!\left( \frac{\pi}{2} - \Bigl( \bigl|\delta_x - \tfrac{\pi}{2}\bigr| + \bigl|\delta_y - \tfrac{\pi}{2}\bigr| \Bigr) \right),
\end{equation}
for ${\delta_x,\delta_y\in [0,\pi]}$, with $\Theta$ the Heaviside step function. The eigenpolarization topology as a function of $\delta_x$ and $\delta_y$ is summarized in the phase diagram plotted in Fig. \figref{fig:fig1}(c). The representative cases considered in Fig. \figref{fig:fig1}(b) are marked by empty diamonds along the black-dashed diagonal line in Fig. \figref{fig:fig1}(c). In the first two cases, the system is topologically trivial (${\nu_\text{s}=0})$ and the eigenpolarization field $\textbf{n}(x,y)$ only covers a portion of the sphere as the position varies across one spatial period. Conversely, for ${\pi/4<\delta<3\pi/4}$, the topology is non-trivial (${\nu_\text{s}=1}$), as observed from the emergence of the skyrmion, and the eigenstates complete a full wrapping around the sphere, as shown in Fig. \figref{fig:fig1}(c).

We note that the same functionality could also be realized within a single nanostructured device, such as a dielectric metasurface or other metamaterial \Cite{capasso1,capasso2,capasso3}.

\begin{figure*}[t!]
    \centering
    \includegraphics[width=\linewidth]{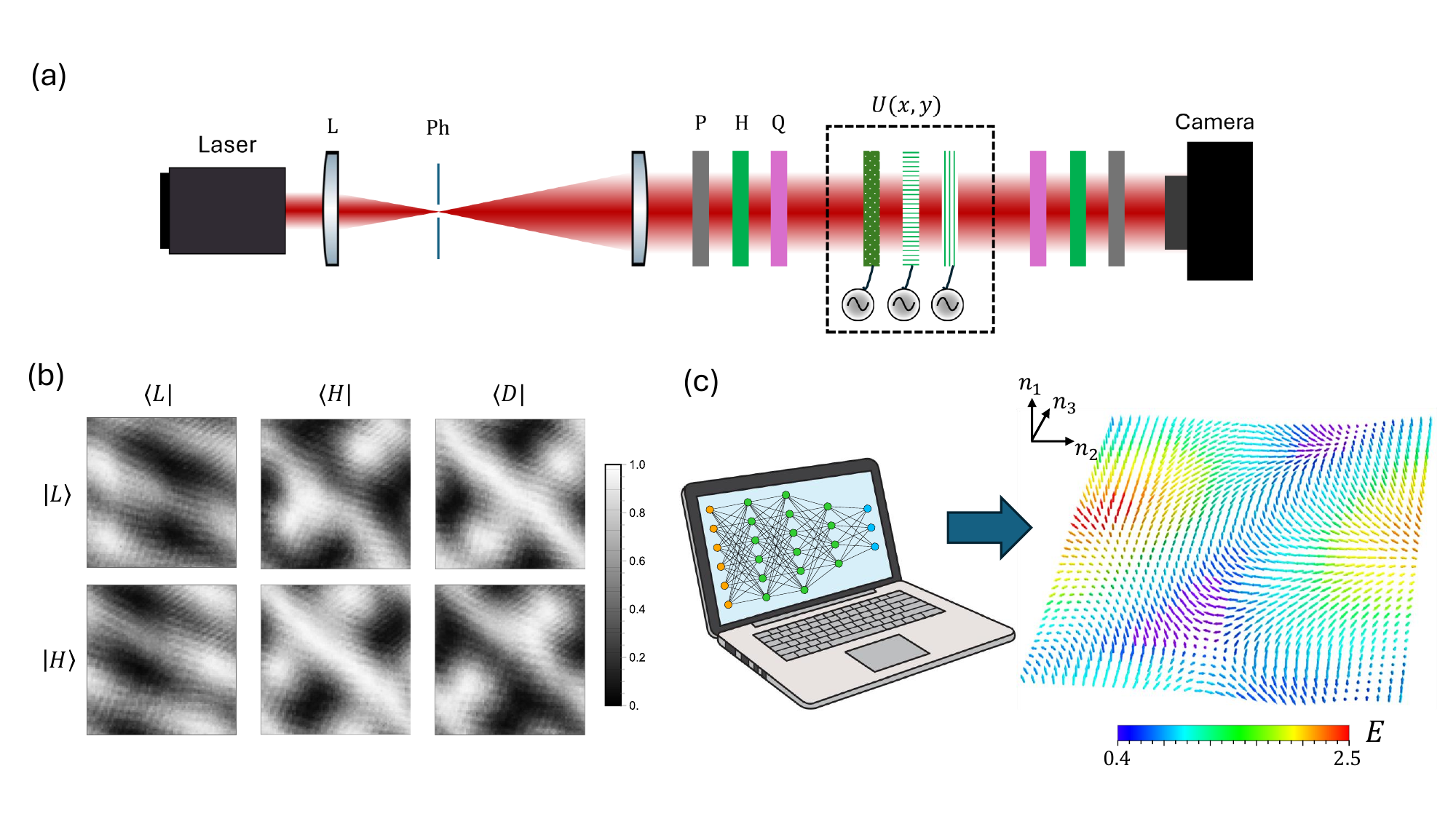}
    \caption{\textbf{Experimental process tomography}. (a) A laser beam is expanded with two lenses (L) and spatially filtered with a pinhole (Ph) placed in the focal plane. Three liquid-crystal metasurfaces implement a space-dependent polarization transformation. The system topology is tuned by applying an AC voltage to each device. Polarimetric measurements are realized by preparing and projecting onto the desired polarization states with a linear polarizer (P), a half-wave plate (H), and a quarter-wave plate (Q). (b) As an example, set of polarimetric images taken for the case ${\delta=\pi/2}$. (c) Such images constitute the input layer of a fully-connected neural network, pretrained to output the model eigenstructure from polarimetric data. The reconstructed patterns can be visualized as arrows overlapping with the local eigenpolarization on the Bloch sphere at each quasi-momentum value. Arrow colors indicate the corresponding energy eigenvalue.}
    \label{fig:fig2}
\end{figure*}


\subsection{Mapping to pseudospin Bloch eigenstates}
\label{sec:bloch}
Light propagation through the periodic devices introduced above gives rise to a diffraction into a discrete set of modes, each carrying quantized transverse momentum ${(m_x,m_y)\Delta k_\perp}$, with ${\Delta k_\perp=2\pi/\Lambda}$ and $(m_x,m_y)$ integer numbers. The complex diffraction network forming in a sequence of such devices can be modeled as a quantum walk \Cite{DErrico:20}, a prototypical example of quantum dynamics on a lattice. Such dynamics takes place in discrete time steps. In the simplest, one-dimensional case, at each step, the walker localized at a given site moves to the left or to the right depending on the state of an internal two-level degree of freedom, referred to as the coin. In our optical analogy, lattice sites are the transverse momentum modes, and the coin is represented by optical polarization \Cite{DErrico:20}. 

In general, quantum walks, and more broadly, tight-binding dynamics with translational invariance, are conveniently described in terms of the variable conjugate to the lattice coordinate, the quasi-momentum. This variable is periodic and defined within a Brillouin zone (BZ). Following the analogy with our optical system, the modes defined above generate a synthetic 2D lattice whose primitive cell in reciprocal space, the BZ, naturally corresponds to one spatial period on the metasurfaces' plane: ${\text{BZ}=[0,\Lambda]^{\otimes2}}$. Accordingly, the quasi-momentum coordinate of the lattice system is physically associated with the transverse position of the beam, ${\textbf{q}\leftrightarrow \textbf{r}}$, with ${\textbf{q}=(q_x,q_y)}$ and ${\textbf{r}=(x,y)}$. The coin (pseudospin) degree of freedom is encoded in the photon polarization.

In this framework, at each transverse position $(x,y)$, $U$ implements a polarization transformation corresponding to the quasi-momentum representation of the unit-step evolution operator, which can be regarded as generated by an effective Bloch Hamiltonian \Cite{PhysRevA.82.033429}:
\begin{equation}
U(x,y):=U(q_x,q_y)=e^{-i H_\text{eff}(q_x,q_y)},
\label{eqn:unitary}
\end{equation}
with ${H_\text{eff}(q_x,q_y)=E(q_x,q_y)\textbf{n}(q_x,q_y)\cdot\bm{\sigma}}$, where ${\bm{\sigma}=\left(\sigma_1,\sigma_2,\sigma_3\right)}$ is the vector of the Pauli matrices, ${\pm E(q_x,q_y)}$ are the quasi-energy values, and ${\textbf{n}=(n_1,n_2,n_3)}$ is a unit vector mapping the pseudospin eigenstate onto a point on the Bloch-Poincaré sphere at each quasi-momentum, as illustrated in Fig. \figref{fig:fig1}(a). Since the evolution is discrete in time, the energy is a periodic quantity. Within this mapping, energy bands are formed by the conjugate phases acquired by the two orthogonal eigenpolarizations at each transverse position (see Eq. \Eqref{eqn:phase-energy}). Since the energy eigenvalues are opposite to each other, we will always target the upper energy band ${E\in[0,\pi]}$ for reference. 

The quantum-walk process simulated via Eq. \Eqref{eqn:opticalU} can be engineered so as to realize a Chern insulator \Cite{DErrico:20}. The specific topological phase is characterized by the Chern number, given by
\begin{equation}
\nu_\text{c}=\frac{1}{2\pi}\int_\text{BZ}\text{d}\textbf{q}\,B_z(\textbf{q}),
\label{eqn:chern}
\end{equation}
where $B_z$ is the Berry curvature:
\begin{equation}
B_z(\textbf{q})=\frac{1}{2}\,\textbf{n}(\textbf{q})\cdot \left(\frac{\partial \textbf{n}}{\partial q_x}\times \frac{\partial \textbf{n}}{\partial q_y} \right).
\label{eqn:berry}
\end{equation}
This establishes the mathematical equivalence between the Skyrme number and the Chern number: ${\nu_\text{s}=\nu_\text{c}\equiv\nu}$ (cf.~Eq. \Eqref{eqn:skyrme} and Eq. \Eqref{eqn:chern}). In particular, the mapping from optical eigenpolarizations to Bloch Hamiltonian eigenstates leaves the topology unchanged, ensuring that the simulated system inherits the same topological structure as the optical setup, as illustrated in Fig. \figref{fig:fig1}(c). 

\begin{figure*}[!th]
    \centering
    \includegraphics[scale=0.51]{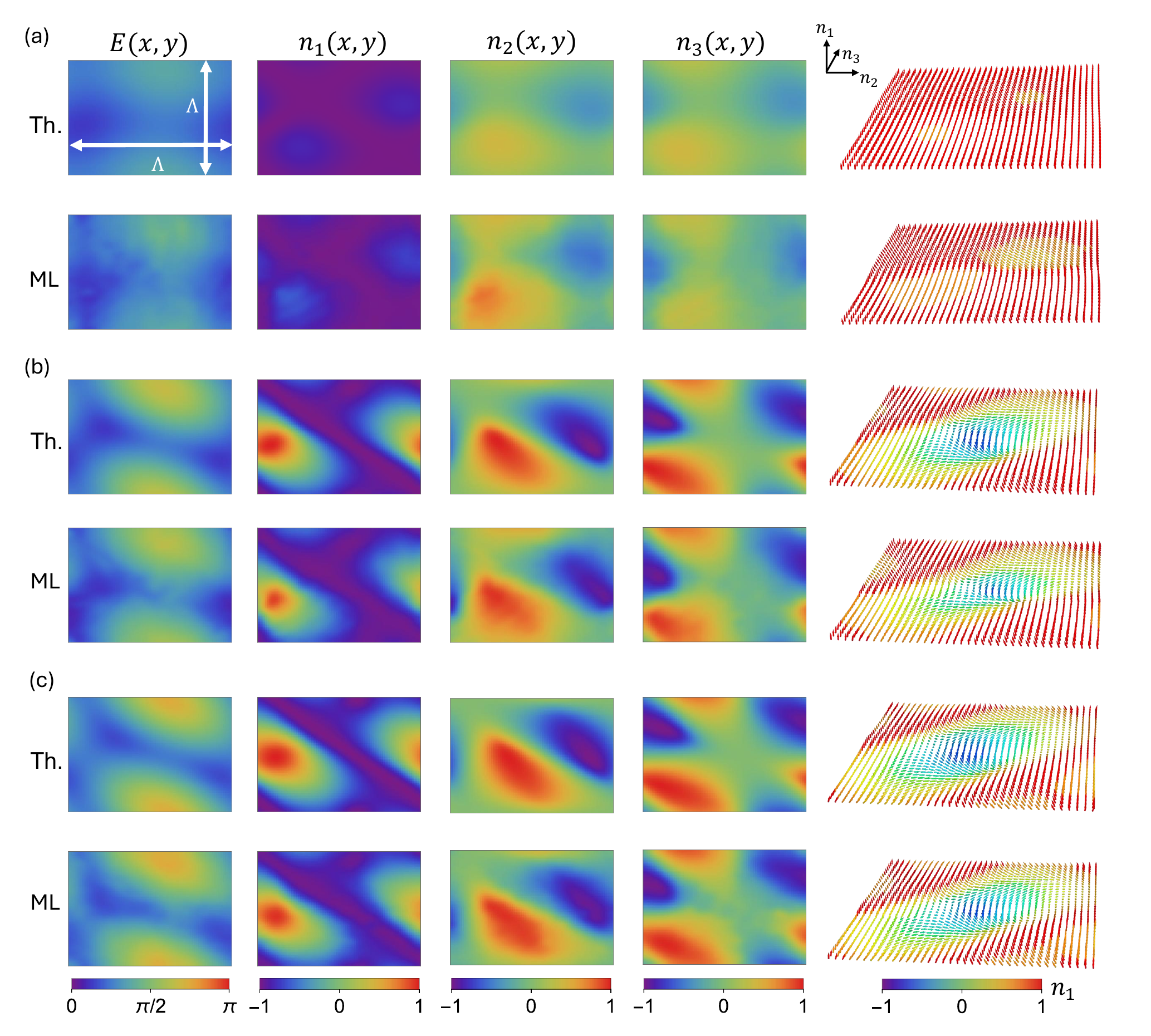}
    \caption{\textbf{Skyrmionic eigenpolarization textures.} Tomographic reconstructions of energy bands $E$ and eigenpolarizations $\textbf{n}$. Experimental patterns obtained via the machine-learning-based tomography (ML) are compared with theoretical (Th.) predictions. The three cases (a) $\delta=\pi/8$, (b) $\delta=5\pi/12$, and (c) $\delta=\pi/2$ are considered. The eigenpolarization pattern is also plotted as arrows, whose orientation gives the direction of the local polarization eigenstate on the Bloch sphere across the BZ. The skyrmionic feature is absent in the topologically trivial phase (a), and only appears in the topologically non-trivial phases (b)-(c). Average fidelities: (a) $(98\pm1)\%$ , (b) $(97\pm1)\%$, (c) $(96\pm3)\%$}
    \label{fig:fig3}
\end{figure*}

\begin{figure*}
\centering
    \includegraphics[scale=0.9]{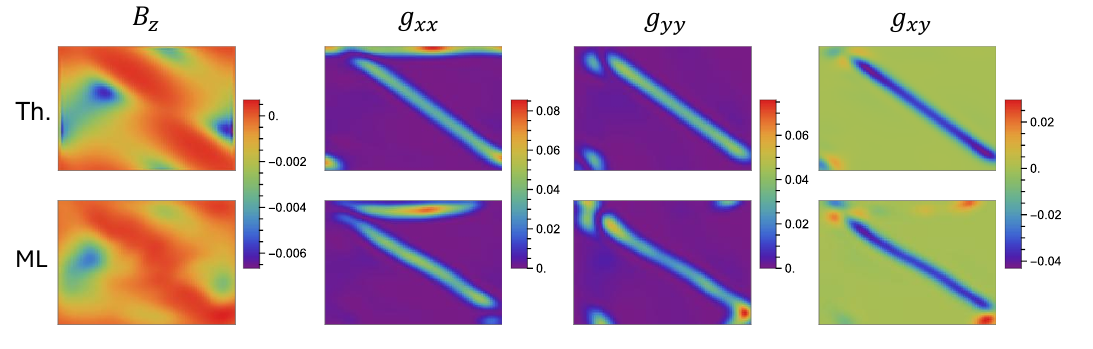}
    \caption{\textbf{Extracting the quantum geometric tensor.} From the reconstructed eigenpolarizations (see Fig. \figref{fig:fig3}), we extract the Berry curvature $B_z$ and the quantum metric components $g_{ij}$, representing the imaginary and real parts of the quantum geometric tensor, respectively. The case ${\delta=\pi/2}$ is shown here. Machine-learning-based (ML) experimental reconstructions are compared with theoretical (Th.) predictions.}
    \label{fig:fig4}
\end{figure*}

\section{Experiment}
\subsection{Process tomography}
To tomographically retrieve the system’s eigenvalues and eigenfunctions, it is convenient to explicitly express $H_\text{eff}$ in Eq. \Eqref{eqn:unitary}:
\begin{equation}
U(x,y)=\cos{E}(x,y)-i\sin{E}(x,y)\left(\textbf{n}(x,y)\cdot\bm{\sigma}\right).
\end{equation}
Our strategy consists of relating a set of polarimetric measurements to the system's eigenstructure. Each polarimetric measurement is realized by setting an input polarization state $\ket{i}$, letting it evolve under $U$, and recording the light intensity after projecting onto $\ket{j}$:
\begin{equation}
I_\text{ij}=I_0\abs{\bra{j}U\ket{i}}^2, 
\end{equation}
where $I_0$ is the total intensity of the light beam. In the following, we will assume $I_0=1$.
In Ref. \Cite{DiColandrea:23}, an optimal set of polarimetric measurements is demonstrated for accurate process tomography of space-dependent SU(2) operators: 
\begin{equation}
\begin{split}
I_{\text{LL}}&=n_3^2 \sin ^2{E}+\cos ^2{E},\\
I_{\text{HH}}&=n_1^2 \sin ^2{E}+\cos ^2{E},\\
I_{\text{LH}}&=\frac{1}{2}\left(1+2 n_1 n_3 \sin^2 {E}+n_2 \sin {2 E}  \right),\\
I_{\text{LD}}&=\frac{1}{2}\left(1+2 n_2 n_3 \sin^2 {E}-n_1 \sin {2 E}  \right),\\
I_{\text{HL}}&=\frac{1}{2}\left(1+2 n_1 n_3 \sin^2 {E}-n_2 \sin {2 E}  \right),\\
I_{\text{HD}}&=\frac{1}{2}\left(1+2 n_1 n_2 \sin^2 {E}+n_3 \sin {2 E}  \right),
\end{split}
\label{eqn:polmeas}
\end{equation}
where ${\ket{H}=(\ket{L}+\ket{R})/\sqrt{2}}$ and ${\ket{D}=\left(\ket{L}+i\ket{R}\right)/\sqrt{2}}$ are horizontal and diagonal polarization states, and we have omitted the explicit dependence of the process parameters ${(E,\textbf{n})}$ on $(x,y)$.

The experimental setup is sketched in Fig. \figref{fig:fig2}(a). A 633-nm laser beam is expanded via a telescopic configuration of two lenses, having focal lengths $f_1=5 $ cm and $f_2=30 $ cm. A 25-$\mu$m pinhole, placed in the focal plane, acts as a spatial filter. The beam is magnified to cover approximately one BZ on the metasurfaces, which corresponds to setting $w_0\simeq\Lambda$, where $w_0$ is the beam waist. Specifically, our fabricated devices feature a spatial period $\Lambda=5$ mm. A linear polarizer (P), a half-wave plate (HWP), and a quarter-wave plate (QWP) are used to prepare the desired input polarization state. The beam then propagates through the three liquid-crystal metasurfaces simulating the quantum walk. Their birefringence is controlled electrically by applying an oscillating field \Cite{Piccirillo2010,Rubano:19}, which allows one to dynamically modify the system topology. The tomographic measurement is completed by projecting the output beam onto a target polarization state with the sequence QWP-HWP-P, and recording the light intensity distribution on a camera placed right after the projection stage to minimize propagation effects. As an example, the set of polarimetric measurements collected for the case ${\delta=\pi/2}$ is shown in Fig. \figref{fig:fig2}(b). Such images are first analyzed by computer and compressed into ${\left(73\times73\right)}$-pixel grids to average over local intensity fluctuations, and then fed into a fully-connected neural network, trained on over $10^6$ examples to associate polarimetric data with the process parameters. A complete reconstruction is obtained in less than 1 s, making machine-learning approaches ideal for real-time monitoring \Cite{Jaouni_2024}. Further details on the network training and hyperparameters can be found in Ref. \Cite{DiColandrea:23}. The eigenpolarization field obtained from the data in Fig. \figref{fig:fig2}(b) is shown in Fig. 
\figref{fig:fig2}(c), where we plot the locally reconstructed pseudospin, ${\textbf{n}(\textbf{q})}$, with the arrow color given by the energy eigenvalue.

\subsection{Reconstruction of eigenpolarizations}
Figure \figref{fig:fig3} shows the reconstructed energy bands and the corresponding Stokes components of the eigenpolarizations across one BZ for the cases (a) ${\delta=\pi/8}$, (b) ${\delta=5\pi/12}$, and (c) ${\delta=\pi/2}$. From the color maps, we observe that the vector field mostly preserves its orientation in the trivial case (see Fig. \figref{fig:fig3}(a)), while a clear inversion appears in the topological non-trivial phase (see Fig. \figref{fig:fig3}(b)-(c)) within the probed region. This topological signature can also be efficiently visualized by plotting the local eigenpolarization texture, where a skyrmionic defect only emerges in the non-trivial phases (b)-(c). In the last column of Fig. \figref{fig:fig3}, the arrows give the orientation of the polarization eigenstates at each quasi-momentum, and their color is associated with the $n_1$ component. A direct comparison with the theoretically expected patterns reveals excellent agreement with the reconstructions obtained from our optimization routine. This is also quantified by the operator fidelity distributions across the region. At each pixel, the operator fidelity is computed as \Cite{PhysRevA.79.012105,Cabrera_2011}
\begin{equation}
F=\dfrac{1}{2}\abs{\text{Tr}\left(U^\dagger_\text{th}U_\text{exp} \right)},
\end{equation}
where $U_\text{th}$ and $U_\text{exp}$ are the theoretical and experimentally reconstructed processes. The average fidelities for the cases considered in Fig. \figref{fig:fig3} are $\bar{F}=(98\pm1)\%,\, (97\pm1)\%, \,(96\pm3)\%$, respectively, where the average is taken over all the pixels and the error is estimated as the standard deviation. By performing the integral of Eq. \Eqref{eqn:skyrme} numerically, we obtain $\nu=0.02,\,1.04,\,0.99$ for the cases $\delta=\pi/8,5\pi/12,\pi/2$, respectively, in perfect agreement with the theoretical predictions (cf.~Fig. \figref{fig:fig1}(c))

From the reconstructed maps, we can also extract local observables associated with the geometry of the parameter space, specifically the Berry curvature and the quantum metric, respectively expressing the geometric-phase difference and the distance between two states in the parameter space \Cite{Provost1980,Berry1989}. The Berry curvature, descending from the imaginary, antisymmetric part of the quantum geometric tensor, acts as an effective magnetic field in momentum space \Cite{RevModPhys.82.1959}, whose orientation is perpendicular to the $(q_x,q_y)$-plane (see Eq. \Eqref{eqn:berry}). The quantum metric represents instead the real, symmetric part of the quantum geometric tensor, whose components are given by \Cite{PhysRevB.97.195422}
\begin{equation}
g_{ij}=\dfrac{1}{4}\left( \dfrac{\partial\vartheta}{\partial q_i}\dfrac{\partial\vartheta}{\partial q_j}+\sin^2{\vartheta}\dfrac{\partial\varphi}{\partial q_i}\dfrac{\partial\varphi}{\partial q_j}\right),
\end{equation}
where $\vartheta=\arccos{n_3}$ and $\varphi=\text{atan2}\left({n_2,n_1}\right)$ are the polar and azimuthal angles of the local eigenpolarization on the Bloch sphere (see Fig. \figref{fig:fig1}(a)). The Berry curvature and the metric components extracted from the reconstruction of Fig. \figref{fig:fig3} for the case $\delta=\pi/2$ are plotted in Fig. \figref{fig:fig4}. Numerical derivatives are computed with a Gaussian filter of ${\sigma=4}$ pixels to suppress high-frequency noise. The extracted observables show very good agreement with theoretical predictions. 

The Skyrme number can also be obtained as the integral of the Berry curvature over the BZ (see Eq. \Eqref{eqn:chern}). A second gauge invariant, referred to as quantum volume, can be analogously extracted by integrating the determinant of the quantum metric over the BZ \Cite{PhysRevResearch.5.L032016,PhysRevB.104.045103,topologygeometry}:
\begin{equation}
\text{Vol}=\dfrac{1}{\pi}\int_\text{BZ}\text{d}^2\textbf{q}\,\sqrt{g_{xx}g_{yy}-g_{xy}^2}.
\end{equation}
This can be intuitively visualized as the area occupied by the eigenpolarizations on the Bloch sphere. Accordingly, a large quantum volume indicates that Bloch states are spread across the BZ, while a small volume suggests that they are less distinguishable as \textbf{q} varies. This argument underlies the inequality ${\text{Vol}\geq \abs{\nu}}$, holding for Chern insulators \Cite{PhysRevB.104.045103,topologygeometry}. For the three cases considered in Fig. \figref{fig:fig3}, we numerically obtain ${\text{Vol}=0.50,\,1.72,\,2.12}$, which proves the inequality is satisfied in all our experimental simulations. 

\begin{figure}
\centering
    \includegraphics[width=0.7\columnwidth]{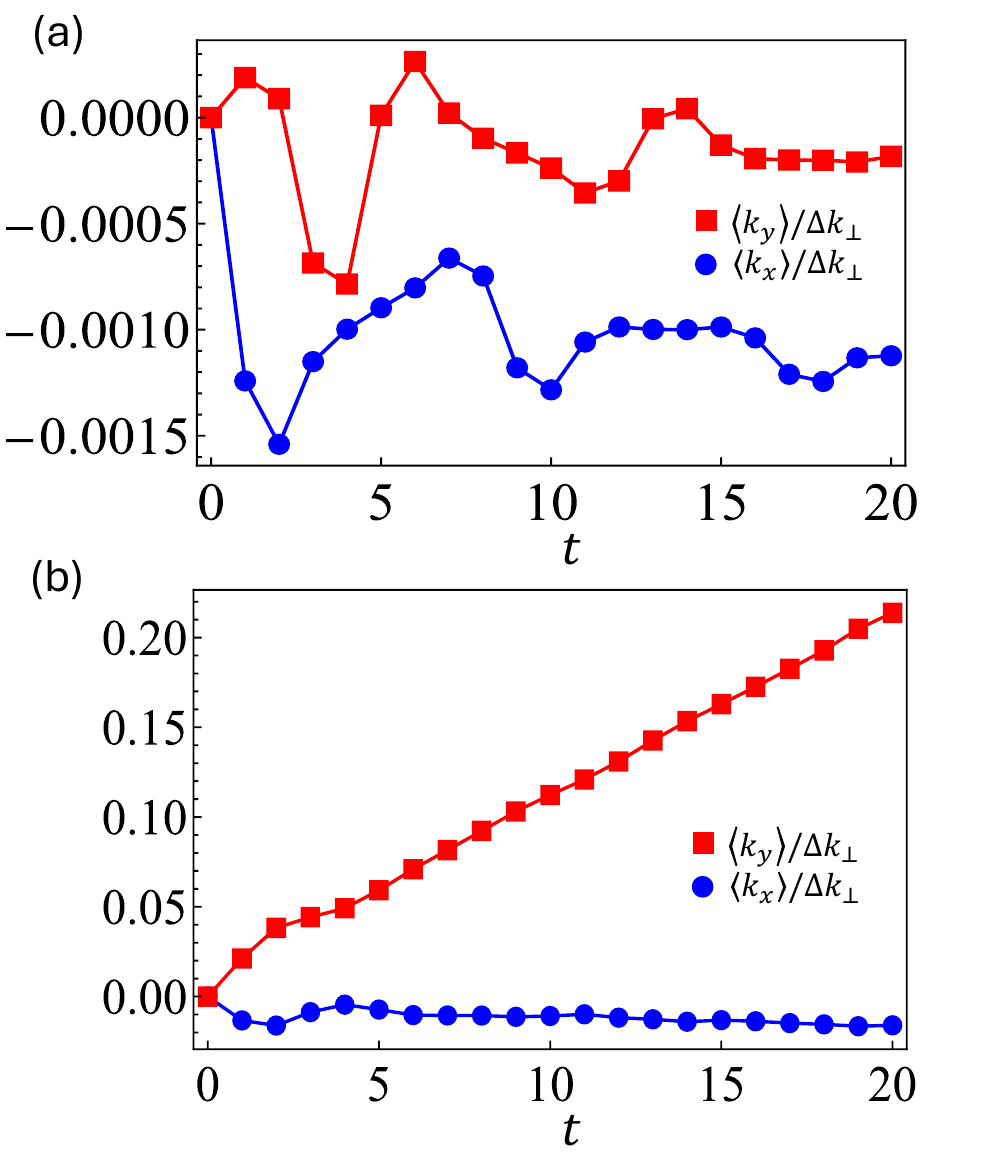}
    \caption{\textbf{Detecting topology through dynamics.} Center-of-mass trajectory under the effect of an external force along the $x$ axis in two simulated experiments at (a) ${\delta=\pi/8}$ and (b) ${\delta=\pi/2}$. In the case of topologically non-trivial bands, a non-vanishing anomalous transverse displacement $\braket{k_y}$ is revealed, as predicted by the adiabatic approximation.}
    \label{fig:fig5}
\end{figure}

\subsection{Dynamical simulation}
One might think that the eigenpolarization topology of the system is just a mathematical curiosity, with no observable physical consequences. However, this is not the case, as we show in the following. Let us assume we cascade $t$ stacks of the three liquid-crystal metasurfaces, each implementing a single step of the operator $U$, as described in Eq. \Eqref{eqn:opticalU}, with the $t$-th $g$-plate along $x$ displaced by an amount given by $t\Delta x$, with $\Delta x=F\Lambda/(2\pi)$. At each transverse position ${(x_0,y_0)}$, we prepare a narrow beam whose polarization is given by the local eigenpolarization ${\ket{\textbf{n}(x_0,y_0)}}$. Its state can be expressed as:
\begin{equation}
\ket{\psi_0(x_0,y_0)}=\mathcal{N}\int\text{d}x\text{d}y\,g_\sigma(x,y)\ket{x_0,y_0}\ket{\textbf{n}(x_0,y_0)},
\end{equation}
where ${\mathcal{N}}$ is a normalization factor and 
\begin{equation}
g_\sigma(x,y)=e^{-\frac{(x-x_0)^2+(y-y_0)^2}{2\sigma^2}}
\end{equation}
is a Gaussian envelope. We let it evolve under the optical sequence, and resolve the transverse-momentum spectrum $(k_x,k_y)$ of the output state by measuring light intensity in the far field. Assume we iterate this procedure over a full spatial period, and then extract the average transverse-momentum components. This is equivalent to simulating the system when initialized in a filled energy band, with equal probability of occupying each state of the band. This setup essentially realizes an optical simulator of the quantum Hall effect, as detailed in Ref. \Cite{DErrico:20}. In the condensed-matter analogue, $F$ plays the role of a constant force acting along $x$. In the adiabatic approximation, i.e., if the applied force is much smaller than the energy bandgap, the beam center of mass will experience an overall displacement along the direction orthogonal to the force proportional to the Chern number \Cite{RevModPhys.82.1959,PhysRevLett.111.135302,PhysRevB.93.245113}:
\begin{equation}
\braket{k_y(t)}\simeq\frac{Ft\nu}{2\pi}\Delta k_\perp.
\label{eqn:semiclassic}
\end{equation}
The results obtained for two simulated experiments at ${\delta=\pi/8}$ and ${\delta=\pi/2}$, spanning 20 time steps, are reported in Fig. \figref{fig:fig5}(a) and (b), respectively. An initial wavepacket with ${\sigma=0.05\Lambda/\pi}$ and a force ${F=\pi/50}$ is considered, and the average is taken over $15\times 15$ discretized transverse-position values. The panels show the average center-of-mass displacement in the far field, i.e., the measured $k_x$ and $k_y$ components (in units of ${\Delta k_\perp}$) at the output. While the total displacement along $x$ remains very close to zero in both phases, an anomalous drift along the $y$ direction is only revealed in the topologically non-trivial case ${\delta=\pi/2}$. Linear fits based on Eq. \Eqref{eqn:semiclassic} reveal $\nu=0.0004\pm0.0009$ and $\nu=1.02\pm0.01$, respectively, in agreement with the expected values.
\vspace{0.5cm}
\section{Conclusion}
We have demonstrated that skyrmionic polarization patterns may emerge in the eigenpolarization texture of structured dielectric materials. This requires the system parameters to be tuned to a configuration which simulates a Chern insulator. Crucially, these optical topologies do not arise from structuring various degrees of freedom of a light beam, as is typically done \Cite{Shen2024,Yang_2025,He2022,Wan2023}, but are instead encoded in the polarization eigenmodes of the material. Consequently, the observed skyrmionic textures constitute intrinsic properties of the material, which only result from its spatial structure and parameter settings. Such topology can also be probed in an anomalous-Hall-effect experimental setup \Cite{DErrico:20}. 

The eigenpolarization patterns have been reconstructed via machine-learning-assisted quantum process tomography, which has proven to outperform standard maximum-likelihood methods in terms of number of measurements, time efficiency, and accuracy \Cite{DiColandrea:23}. Experimentally, this has been accomplished by probing the system under different polarimetric settings, which also allowed us to extract relevant properties of the associated Hilbert-space manifold of the Bloch pseudospin eigenstates in post-processing. 

An interesting prospect could be to train a neural network to infer
the topological invariant directly from experimental data, for instance, by combining partial tomographic information collected in reciprocal spaces \Cite{FQPT}. Such a routine could also be trained to directly retrieve the components of the quantum metric. Since measurements of the quantum geometric tensor from real-space wavepacket displacements proved extremely challenging \Cite{Gianfrate2020}, tomographic approaches may provide a practical and viable alternative. 
Further directions include the generalization of these methods to non-Hermitian systems \Cite{ashida2020non,RevModPhys.93.015005,NHgeometry, 10.1063/5.0274332}, as well as to multi-band \Cite{PhysRevB.74.024408} and higher-order Chern insulators \Cite{hoTI}. 

\bibliography{bibliography}

\bigskip
\noindent\textbf{Acknowledgments.}
This work was supported by the PNRR MUR project PE0000023-NQSTI.
\bigskip
\newline \noindent \textbf{Disclosures.}
The authors declare no conflicts of interest.
\end{document}